\documentclass[traditabstract,longauth]{aa}  
\makeatletter
\def\@LN#1#2{} 
\def\@LN@col#1{} 
\let\linenumbers\relax
\let\nolinenumbers\relax
\makeatother

\usepackage[T1]{fontenc}
\usepackage{txfonts}
\usepackage{graphicx}
\usepackage[dvipsnames]{xcolor}
\usepackage{amsmath,amssymb}
\usepackage{lscape}
\usepackage{subcaption}
\usepackage{placeins}
\usepackage{hyperref}
\hypersetup{
    colorlinks=true,
    citecolor={blue}
}

\usepackage{natbib}
\bibpunct{(}{)}{;}{a}{}{,} 

\titlerunning{Neutron star heating vs. HST observations}
\authorrunning{Rodríguez et al.}

\begin{document} 

\title{Contrasting neutron star heating mechanisms with \emph{Hubble Space Telescope} observations}

\author{
Luis E. Rodr\'iguez\thanks{e-mail: lsrodrig@uc.cl}\inst{1}
\and Andreas Reisenegger\inst{2,3}
\and Denis Gonz\'alez-Caniulef\inst{4}
\and Crist\'obal Petrovich\inst{1,5}
\and George Pavlov\inst{6}
\and S\'ebastien Guillot\inst{4}
\and Oleg Kargaltsev\inst{7}
\and Blagoy Rangelov\inst{8}
}

\institute{
Instituto de Astrof\'isica, Pontificia Universidad Cat\'olica de Chile, Av. Vicu\~na Mackenna 4860, Macul, Santiago, Chile
\and Departamento de F\'isica, Facultad de Ciencias B\'asicas, Universidad Metropolitana de Ciencias de la Educaci\'on, Av. Jos\'e Pedro Alessandri 774, \~Nu\~noa, Santiago, Chile
\and {Centro de Desarrollo de Investigaci\'on UMCE, Universidad Metropolitana de Ciencias de la Educaci\'on, Santiago, Chile}
\and Institut de Recherche en Astrophysique et Planétologie (IRAP), UPS-OMP, CNRS, CNES, 9 avenue du Colonel Roche, BP 44346, F-31028 Toulouse Cedex 4, France
\and Department of Astronomy, Indiana University, Bloomington, IN 47405, USA
\and Department of Astronomy \& Astrophysics, Pennsylvania State University, 525 Davey Lab., University Park, PA 16802, USA
\and Department of Physics, The George Washington University, 725 21st Street NW, Washington, DC 20052, USA
\and Department of Physics, Texas State University, 601 University Drive, San Marcos, TX 78666, USA
}

\date{Received XX Month 2025 / Accepted XX Month 2025}


\abstract{Passively cooling neutron stars (NSs) are expected to reach undetectably low surface temperatures $T_s < 10^4$~K within less than $10^7$~yr. However, \emph{Hubble Space Telescope} (HST) observations have revealed likely thermal ultraviolet emission from the Gyr-old millisecond pulsars PSR~J0437$-$4715 and PSR~J2124$-$3358 and the $\sim 10^{7-8}$ yr-old classical pulsars PSR~B0950$+$08 and PSR~J0108$-$1431, implying temperatures $T_s \sim 10^5$~K and thus suggesting the presence of some heating mechanism.  
Here, the thermal evolution {for each of these} NSs was computed, considering rotochemical heating in the NS core with normal or Cooper-paired matter, 
vortex creep in the inner crust, and crustal heating through nuclear processes. The results were contrasted with the observations,  
including also the stringent upper limit on the temperature of  
PSR~J2144$-$3933.  
None of the heating mechanisms by itself was found to account for all the observations. The high temperature of PSR~J0437$-$4715 can be explained by rotochemical heating in the presence of a large Cooper pairing gap $\Delta_i\sim 1.5\,\mathrm{MeV}$ for either neutrons or protons (or an equivalent combination of both), but this mechanism would require an unrealistically short  initial rotation period $P_0\lesssim 1.8\,\mathrm{ms}$ to account for the high temperature of the classical pulsar PSR~B0950$+$08. Conversely, the latter can be explained by rotochemical heating with modified Urca reactions in normal matter or by vortex creep with an excess angular momentum parameter $J\sim 3\times 10^{43}\mathrm{erg\,s}$, but these models are insufficient to account for the former. However, a model that includes rotochemical heating with a large Cooper pairing gap together with vortex creep is consistent with the temperature measurements of these two pulsars as well as the upper limits for the other three. Moreover, it predicts that the temperatures of the latter three pulsars should be close to these upper limits, suggesting that deeper observations and/or a wider wavelength coverage for these should yield a strong probe of this model.
}

\keywords{
dense matter -- equation of state -- stars: neutron -- pulsars: individual: PSR~J0437$-$4715, PSR~J2124$-$3358, PSR~B0950$+$08, PSR~J0108$-$1431, PSR~J2144$-$3933 -- ultraviolet: stars
}

\maketitle


\section{Introduction}\label{sec:intro}
From both the theoretical and experimental point of view, the properties of matter at very high densities, above the nuclear saturation density $\rho_0 = 2.8\times10^{14}$ g cm$^{3}$, are still unknown and remain as an open problem for fundamental physics.  However, they can be studied through the thermal evolution of neutron stars (NSs), since the matter in their interior can reach several times $\rho_0$. By contrasting thermal evolution models with temperature measurements of NSs of different ages, we can learn, for example, about the composition and potential superfluid and superconducting states of very dense matter. 

According to standard cooling models, NSs are born with a very high interior temperature $\sim 10^{11}$ K and rapidly cool by emitting thermal neutrinos produced in their core. Neutrinos act as the main cooling agent and control the internal temperature for the first $\sim 10^5$ yr.  In fact, several X-ray observations of young NSs have been made to contrast their temperatures with cooling curves and study the neutrino emission process, as well as the role of the composition, mass and state of matter of the NS \citep[for a review see, e.~g.,][]{Yako2004}.

Beyond the neutrino cooling age, photon emission from the NS surface becomes the main cooling mechanism. By $\sim10^{7}~\mathrm{yr}$, the temperature of passively cooling NSs is expected to drop below $\sim10^{4}$ K, becoming undetectable by current telescopes. However, observations with the Hubble Space Telescope (HST) have revealed plausibly thermal ultraviolet (UV) radiation from:
\begin{itemize}
    \item millisecond pulsars (MSPs) PSR~J0437$-$4715 (hereafter J0437; \citealt{Kar2004,Durant2012}) and PSR~J2124$-$3358 (hereafter J2124; \citealt{Range2017}), with characteristic ages $\tau\equiv P/(2\dot P)>10^9\,\mathrm{yr}$, where $P$ is the rotation period and dots denote time derivatives, and 
    \item classical pulsars (CPs) PSR~B0950+08 (hereafter B0950; \citealt{Pavlov2017,Abramkin2022}) and  PSR~J0108$-$1431 (hereafter J0108; \citealt{Abramkin2021}), with $\tau>10^7\,\mathrm{yr}$. 
\end{itemize}
Thus, unless the true ages of these pulsars are much smaller than their characteristic ages (which is very unlikely), this suggests the presence of internal heating mechanisms operating in very old NSs. Additionally, the non-detection of the CP PSR~J2144$-$3933 (hereafter J2144; \citealt{Guillot2019}), which makes it the coolest NS known so far, also provides a constraint on potential heating mechanisms.

\citet{GonRei2010} analyzed several proposed NS heating mechanisms, namely rotochemical heating \citep{Reis1995}, vortex creep \citep{Alpar1984}, dark matter accretion \citep{Kou10,deLava}, crust cracking \citep{Baym71}, and magnetic field decay \citep{Gold92,Pons25}, finding that only rotochemical heating and vortex creep can plausibly explain the inferred surface temperature of J0437, the only pulsar mentioned above that had been detected by HST at that time (see also \citealt{Fujiwara2024DM}).

For the younger, more strongly magnetized CPs, magnetic field dissipation might also give an important contribution. However, models for this process are not yet fully developed (e.~g., \citealt{
Pons25}, and references therein), as it depends on many uncertain assumptions, such as the initial structure of the magnetic field (e.~g., \citealt{Becerra2022,Dehman2023}), the impurity content of the crust \citep{Cumming2004}, and superfluid/superconducting properties of the core \citep{Gusakov2020b}. Therefore, we will not include it in the present study.

Another potential heat source we do not consider here are relativistic particles from the magnetosphere hitting the NS surface (e.~g., \citealt{Harding2002}). As discussed by \citet{Kar2004}, it is unlikely that a substantial fraction of this heat gets distributed over the whole NS surface in order to contribute significantly to the UV emission.

One of the processes we do consider is vortex creep {\citep{Alpar1984}}. It is based on the idea that neutrons can become superfluid in the deep layers of the NS crust, forming quantized vortices that dissipate heat by friction with the crustal nuclear lattice as they move outward as a result of the NS spin-down. {For recent work on this process, see \citet{Fujiwara2024VC,Nam2025}.}

Rotochemical heating is produced by non-equilibrium beta reactions in the NS core due to the departure from beta equilibrium produced by the continuous spin-down of the NS. \citet{GonRei2010} considered the case of rotochemical heating in NS cores with normal (non-Cooper-paired) matter, as in the study of \cite{Fer2005}. However, important theoretical progress has been made afterwards. \citet{Petro2010} modeled rotochemical heating in the presence of Cooper pairing, that is, with superfluid neutrons and/or superconducting protons, considering a uniform {and isotropic} Cooper pairing gap. The latter suppresses reactions, allowing chemical imbalances to grow more than in the case of normal matter, which eventually leads to a higher heat input for old, fast rotating NSs \citep{Reis1997}. \citet{Gonj2015} extended this study to the case of a non-uniform and anisotropic energy gap, finding that the evolution is similar to the case with a uniform gap equal to its minimum value. {\citet{Yanagi2020} made it even more general by considering non-uniform gaps for both neutrons and protons.} More recently, \citet{Kantor2021} took into account that the proton superconductor in the NS core can be threaded by a magnetic field, either through quantized flux tubes with normal cores (type-II superconductor) or through normal domains of uncertain geometry (type-I superconductor). This allows a higher rate of Urca reactions within these regions, as previously suggested by \citet{Schaab1998}.

Furthermore, \citet{Gus2015} proposed a variant of rotochemical heating, in which pulsars that have previously accreted a substantial fraction of their crust can be heated by pycnonuclear reactions in the deep layers of the latter. The continuous NS spin-down increases the departures from nuclear equilibrium, triggering nuclear reactions as the density increases. This mechanism operates only in MSPs and, as presented in the original paper, would produce luminosities comparable to those observed in J0437 and J2124. Recent insights into the equation of state of accreted crusts \citep{Gusakov2020a} are likely to strongly reduce this effect. Since this has not yet been calculated, we will use the original prediction as an upper limit.

The goal of this work is to revisit the problem of heating mechanisms in old NSs, calculating their thermal evolution curves with vortex creep, crustal heating, and rotochemical heating with and without Cooper pairing ({assuming isotropic gaps and} neglecting the effect of magnetic flux penetration in the proton superconductor) and comparing them to the HST observations mentioned above.

Sect. \ref{sec:observational} describes the observational data available on the five old pulsars used in this study. Sect.~\ref{sec:thermal_evolution} gives a general discussion of the thermal evolution of NSs, whereas Sect.~\ref{sec:heating} discusses the three internal heating mechanisms considered in the present work. Sect.~\ref{sec:simulations} discusses the numerical simulations and their results and compares them to the observational data, and Sect.~\ref{sec:conclusions} presents our conclusions. {In Appendix~\ref{appendix}, we analyze thermal oscillations that occur when the energy gap is large and uniform and their suppression in the case of reduced reaction rates.}

\section{Observational data}\label{sec:observational}

We consider surface temperature measurements derived from HST observations of old CPs and  MSPs, which are summarized in Table~\ref{table:summary}. In general, the thermal emission from NSs is expected to be reprocessed by an atmosphere. However, for the relatively large magnetic fields ($B\sim10^{12}\,\mathrm{G}$) and cool surface temperatures ($T_s\sim10^5\,\mathrm{K}$) of old CPs, atmosphere models are rather uncertain (even a condensed surface might be present; see, e.~g., \citealt{Potekhin2014}). Nevertheless, state-of-the-art hydrogen atmosphere models generally follow a trend in which, for increasingly strong magnetic fields, the UV tail of the spectrum at a given effective temperature approaches the Rayleigh–Jeans tail of a blackbody spectrum of the same temperature \citep[e.~g.][]{Lloyd2003}\footnote{Although more uncertain, for some models of condensed surface emission, the UV spectrum remains well below the Rayleigh-Jeans tail; see \citet{Potekhin2014}.}. Consistently, and in order to simplify the analysis of HST observations, we consider surface temperatures of CPs derived from blackbody spectral fits. 

We reanalyse and update the upper limit on the surface temperature of J2144 accounting for the latest recalibrations of the HST's Solar Blind Channel (SBC)  instrument \citep{Avila2019} in imaging mode as explained on the Space Telescope Science Institute webpage\footnote{\url{https://www.stsci.edu/hst/instrumentation/acs/data-analysis/zeropoints}}. Following \citet{Guillot2019}, we performed blackbody fits to the photometric UV data, accounting for the uncertainties in the distance, dust extinction, and NS radius, 
and we consider the resulting temperature as an upper limit, since we did not include the potential contribution of a non-thermal component from the pulsar magnetosphere. The temperatures for B0950 and J0108 are used as reported by \citet{Abramkin2022} and \citet{Abramkin2021}, respectively, who also fitted blackbody models and included the ACS/SBC recalibrations.

\begin{table}
\caption{Properties of the pulsars considered in the present paper.} 
\begin{center}
\begin{tabular}[H]{lcccc}
\hline

Pulsar & 
$P$ [ms]       & 
log $B$ [G] & 
log  $\tau$ [yr] &
$T^\infty_s/(10^5 \mathrm{K})$\\
\hline

J0437$-$4715  & 
5.75          & 
8.45          & 
9.82          & 
$2.36^{+0.15}_{-0.14}$\\ 

J2124$-$3358 & 
4.93         &
8.28         &
10.02        & 
$<4.9$ 
\\
\hline

B0950$+$08 & 
253      & 
11.38    & 
7.24     & 
$0.6-1.2$\\ 

J0108$-$1431 & 
808      & 
11.36    & 
8.29     & 
$<0.6$\\ 

J2144$-$3933 & 
8510         &
12.27        &
8.43         & 
$<0.3$\\ 
\hline
\end{tabular}
\tablefoot{The columns contain the name of the pulsar, its rotation period ($P$), magnetic field strength ($B$) inferred from spin-down, characteristic age ($\tau$), and measured surface redshifted temperature ($T_s^\infty$). The values of $P$, $B$, and $\tau$ were taken from the ATNF Pulsar Catalogue (\url{https://www.atnf.csiro.au/research/pulsar/psrcat/}; \citealt{Manchester2005}), with $B$ and $\tau$ corrected for the ``Shklovskii effect'' \citep{Shklovskii1970,Camilo1994}. The values of $T_s^\infty$ were taken from the HST observations of 
J2124 \citep{Range2017}, B0950 \citep{Abramkin2022}, J0108 \citep{Abramkin2021}, and J2144 \citep{Guillot2019}, applying recalibrations for J2124 and J2144, 
and from a joint analysis of HST and ROSAT observations for J0437 (see Sect.~\ref{sec:observational} for details).}   \label{table:summary}
\end{center}
\end{table}

In the case of MSPs, their relatively low magnetic field, $B\sim10^8\,\mathrm{G}$, greatly simplifies the modeling of the atmospheric thermal emission \citep[see, e.~g.,][]{Zavlin1996}. For J0437, we consider spectral fits of non-magnetic hydrogen atmosphere models to the cold thermal component as in the recent analysis of \cite{Gon2019}, but using UV spectroscopic data taken with the Space Telescope Imaging Spectrograph (STIS; \citealt{Kar2004}) instead of the SBC instrument \citep{Durant2012}, as the latter requires recalibration that is not yet available for the spectroscopic mode\footnote{According to the HST Help Desk.}. 
We also include the soft X-ray emission of J0437 as observed by ROSAT \citep[see also][]{Guillot2016}. Including a Gaussian prior on the extinction and accounting for the effect of the hot polar caps, as described in \citet{Gon2019}, we obtain the best fitting parameter values $T^\infty_s=2.36^{+0.15}_{-0.14}\times 10^5\,\mathrm{K}$, $R=13.00^{+1.16}_{-0.96}\,\mathrm{km}$, $N_H=(1.4\pm 0.3)\times 10^{20}\,\mathrm{cm}^{-2}$, and $E(B-V)=0.01\pm 0.01$, very similar to those reported in that paper. The resulting surface temperature is more tightly constrained than those obtained by \citet{Kar2004,Durant2012}, and \citet{Bogdanov2019}, which were used in previous comparisons with theoretical models (such as \citealt{Fer2005,GonRei2010,Petro2010,Gonj2015,Kantor2021}). We note that \citet{Kar2004} and \citet{Durant2012} assumed blackbody models to infer the temperature, which do not appear to be consistent with realistic extinction values and are known to yield lower temperatures than hydrogen (or helium) atmosphere models \citep{Gon2019}. In addition, the NICER data analysis of \citet{Bogdanov2019} used a very crude background model, which could explain the difference in temperature deduced from the spectral analysis. Therefore, our new result might be more trustworthy than previous ones.

We also note that J0437 is the only pulsar in our sample that is in a binary system, which has allowed for a precise measurement of its mass, $M=1.418\pm 0.044\,M_\odot$ \citep{Reardon2024}, through radio timing, and its coordinate radius has been measured from its NICER X-ray lightcurve and spectrum to be $R=11.36^{+0.95}_{-0.63}\,\mathrm{km}$ \citep{Choudhury2024}, consistent with the value given above (see \citealt{Hoogkamer2025} for a different analysis). Furthermore, the temperature of the white-dwarf companion implies an age $6.0\pm 0.5\,\mathrm{Gyr}$ \citep{Durant2012}, close to the pulsar's characteristic age, $\tau=6.6\,\mathrm{Gyr}$, which means that, at the end of the mass transfer stage, the pulsar must have been spinning much faster than at present. Assuming dipole spin-down with a constant magnetic dipole moment (braking index $n=3$), the initial period must have been in the range $0.7<P_0/{\mathrm{ms}}<2.4$ (at $68\%$ confidence), compared to the current period $P=5.75\,\mathrm{ms}$. 

For J2124, the available data do not yet allow to distinguish what fraction of the UV emission is thermal vs. a non-thermal power law, which only makes it possible to establish an upper limit on the surface temperature. Following \cite{Range2017}, we perform a blackbody fit to the recalibrated UV photometric data, accounting for the uncertainties associated to the distance and extinction and yielding a model-dependent upper limit $T^\infty_{s,BB}<1.8\times 10^5\,\mathrm{K}$. To this result, we apply the conversion formula to a hydrogen atmosphere provided by \citet{Gon2019} at the end of their Sect. 3, which yields an upward correction by a factor 2.7, and thus an absolute upper limit $T^\infty_s<4.9\times 10^5\,\mathrm{K}$.

\section{Thermal evolution models}
\label{sec:thermal_evolution}

The two classes of NSs studied in this paper (CPs and MSPs) have substantially different properties and evolutionary histories. CPs are born in the core collapse of a massive star, which leads to a supernova explosion and a remaining proto-NS with a very high initial core temperature $T_c\sim 10^{11}~\mathrm{K}$, a strong external magnetic dipole field $B\sim 10^{11-13}~\mathrm{G}$ (in addition to the likely presence of higher multipoles), and a moderate initial rotation period, $P_0\sim 10~\mathrm{ms}-0.5~\mathrm{s}$ (e.~g., \citealt{Faucher-Giguere2006,InPerA,Du2024}). The magnetic dipole torque causes the star to progressively spin down, while the cooling and heating processes studied in the present paper cause the temperature to evolve over time. We will assume that the magnetic field is constant in time, ignoring the possibility of a substantial evolution during this phase (e.~g., \citealt{Gold92}).

If a NS is in a binary system, it might accrete $\sim 0.1\,M_\odot$ of matter \citep{Ozel2012,Antoniadis2016}, heating it up to a core temperature $T_c\sim 10^{8-9}~\mathrm{K}$ \citep{Nava-Callejas2025} and reducing its rotation period to $\sim 1-10~\mathrm{ms}$ \citep{Sautron2025}, thus turning it into a MSP. Either the accretion process itself \citep{Bisnovatyi-Kogan1974} or the long intervening time \citep{Cruces2019} also leads to a reduction of the magnetic dipole field to $B\sim 10^{8-9}\mathrm{G}$, making the spin-down of MSPs much slower than that of CPs. In what follows, we consider the ``birth'' or initial state of a MSP to be the time when accretion stops and the NS evolution is no longer affected by its companion (which might by then have disappeared).

The thermal evolution of a NS depends on the reactions and the state of matter in the core, whose outer part is composed of degenerate neutrons ($n$), protons ($p$), electrons ($e$), and muons ($\mu$). The composition of the inner core is uncertain, so we will assume the whole core to have this composition. The redshifted core temperature, $T_c^\infty$, can be taken to be uniform because the typical thermal evolution time-scales are much longer than the diffusion time \citep{Reis1995}. Its evolution can be computed from \citep{Thorne1977}:
\begin{equation} \label{ThermalEvol}
\dot T_c^\infty = \frac{1}{C}\Big(L_{H}^{\infty}-L_{\nu}^{\infty}-L_{\gamma}^{\infty}\Big),
\end{equation}
where $C$ is the total heat capacity of the star, $L_{H}^{\infty}$ is the total power injected by heating mechanisms, $L_{\nu}^{\infty}$ is the power emitted as thermal neutrinos and antineutrinos from the NS interior, and
\begin{equation} \label{photon luminosity}
    L_{\gamma}^{\infty}=4\pi R_\infty^2\sigma (T^\infty_s)^4
\end{equation} 
is the power released as thermal photons from the surface. In the latter equation, $\sigma$ is the Stefan-Boltzmann constant, $R_\infty$ is the NS radius, and $T_s^\infty$ is its (effective) surface temperature. Throughout this paper, dots denote time-derivatives, and the superscript or subscript $\infty$ labels 
quantities redshifted to infinity, i.~e., as measured 
far away from the NS. The 
temperatures of the core, $T_c^\infty$, and the surface, $T_s^\infty$, have a one-to-one relation to each other, which depends on the envelope composition \citep{Gud83,Potekhin1997}. However, in the late-time evolution studied in this work, both $T_c^\infty$ and $T_s^\infty$ are low, which implies that the cooling in chemical equilibrium is strongly dominated by photon emission from the surface, which depends only on $T_s^\infty$. Also, at late times, the heat released in the NS core can be considered to be emitted instantaneously,\footnote{We keep the term $L_{\nu}^{\infty}$ on the right-hand side because for rotochemical heating it is closely connected and usually non-negligible compared to $L_{H}^{\infty}$. Both become temperature-independent for cool NSs.} 
\begin{equation}\label{heating=cooling}
    L_{\gamma}^{\infty}\approx L_{H}^{\infty}-L_{\nu}^{\infty},
\end{equation}
so the thermal evolution becomes nearly independent of the NS's heat capacity and core temperature. Thus, the envelope model, although required as an input for the simulations, does not affect the late-time results for $T_s^\infty$. 

Two main kinds of weak interaction processes can contribute to the neutrino luminosity, $L_\nu^\infty$, and internal heating, $L_H^\infty$, namely direct Urca (Durca) reactions,
\begin{equation}\label{dur}
n \rightarrow p+\ell+\bar\nu_\ell,\qquad p +\ell \rightarrow n+ \nu_\ell,
\end{equation}
and modified Urca (Murca) reactions,
\begin{equation}\label{mur}
n + N \rightarrow p+\ell+\bar\nu_\ell + N,\qquad p +\ell + N \rightarrow n+ \nu_\ell + N,
\end{equation}
where $\ell$ stands for a negatively charged lepton (electron or muon), and $\nu_\ell$ and $\bar\nu_\ell$ for the corresponding neutrino or antineutrino. The latter reactions differ from the former by the presence of a spectator nucleon $N$ (a neutron or proton), which does not change flavor, but is needed to satisfy the conservation of energy and momentum in the case of low abundance (and therefore small Fermi momentum) of protons and leptons. This allows the Murca reactions to be present in any NS core, whereas the much faster Durca reactions {are only allowed above a certain threshold density.} 
Thus, the latter are likely to be present only in the higher-density regions of relatively massive NSs. The luminosities due to both Durca and Murca reactions for normal matter in chemical equilibrium can be written as
\begin{equation} \label{eq:normalUrcaLum}
L_{\nu,\mathrm{eq}}^\infty=(\tilde L_e+\tilde L_\mu)(T_c^\infty)^q,
\end{equation}
where $q=6$ for Durca and $q=8$ for Murca reactions \citep{Yako2001}. The constants $\tilde L_\ell$ correspond to reactions involving electrons ($\ell=e$) and muons ($\ell=\mu$) and are given in Table~\ref{table:parameters} for the particular models used in the present paper. For matter out of chemical equilibrium (discussed in Sect.~\ref{sec:rotochemical}) and/or with Cooper pairing, these luminosities must be multiplied by reduction factors that depend on the chemical imbalances and/or pairing gaps, respectively (e.~g., \citealt{Petro2010} and references therein).

Both CPs and MSPs start their evolution at a high temperature, so their cooling is initially dominated by neutrino emission ($L_\nu^\infty$), with photon emission ($L_\gamma^\infty$) taking over $\sim 10^5\mathrm{yr}$ later, at lower temperatures. The heating mechanisms contributing to $L_{H}^{\infty}$, which become relevant at relatively low temperatures, are explained in Sect.~\ref{sec:heating}.

\section{Heating mechanisms}\label{sec:heating}

\subsection{Rotochemical heating}\label{sec:rotochemical}

\begin{table}
\caption{Parameter values calculated for the two NS models used in this study.} 
\begin{center}
\begin{tabular}[H]{lcccc}
\hline

Variable &  Units & Model M & Model D  \\
\hline
$M$          & $M_\odot$                            & $1.44$      & $1.76$ \\
$R$          & $\mathrm{km}$                        & $11.5$     & $11.8$ \\
$R_\infty$   & $\mathrm{km}$                        & $14.4$      & $15.8$ \\
\hline
$\tilde L_e$        & $\mathrm{erg\,s^{-1}\,K}^{-q}$       & $2.69\times 10^{-32}$ &  $6.18\times 10^{-8}$\\
$\tilde L_\mu$      & $\mathrm{erg\,s^{-1}\,K}^{-q}$       & $1.82\times 10^{-32}$ &  $6.53\times 10^{-8}$\\
\hline
$W_{npe}$    & $10^{-13}\,\mathrm{erg\,s}^2$                  & $-1.70$ &  $-1.84$ \\ 
$W_{np\mu}$  & $10^{-13}\,\mathrm{erg\,s}^2$                  & $-2.22$ &  $-2.14$ \\ 
$Z_{np}$     & $10^{-61}\,\mathrm{erg}$                       & $3.04$ &   $2.70$      \\
$Z_{npe}$    & $10^{-61}\,\mathrm{erg}$                       & $8.70$ &  $7.52$ \\
$Z_{np\mu}$  & $10^{-61}\,\mathrm{erg}$                       & $10.7$ &  $9.02$ \\
$K_e$        & $10^{47}\,\mathrm{s}^2$                       & $2.74$ & $3.58$ \\
$K_\mu$      & $10^{47}\,\mathrm{s}^2$                       & $3.34$ &  $3.69$      \\
$K$          & $10^{47}\,\mathrm{s}^2$                       & $6.09$ &  $7.27$ \\

\hline
\end{tabular}
\tablefoot{Model M, with the equation of state A18+$\delta v$+UIX* \citep{Akmal1998}, allows only Murca reactions (for which the exponent $q=8$), whereas Model D is based on the equation of state BPAL32 \citep{Prakash1988} and allows Durca reactions ($q=6$). The first three parameters in the list are the NS mass $M$, the coordinate radius $R$, and the redshifted radius $R_\infty$; the remaining parameters are defined by the equations of Sect.~\ref{sec:rotochemical}.
}\label{table:parameters}
\end{center}
\end{table}

Since the birth of a NS, its continuous spin-down reduces the centrifugal force, increasing the local pressure and density {everywhere inside the star}. This changes the Fermi energies of the different particle species in the core, producing chemical imbalances,
\begin{equation}\label{imbalances}
    \eta_\ell\equiv\mu_n-\mu_p-\mu_\ell,
\end{equation}  
where $\mu_i$ are the chemical potentials of each particle species (labeled as in Eqs.~\ref{dur} and \ref{mur}), and inducing non-equilibrium Urca reactions that act in the sense of restoring the chemical equilibrium. As for the temperature, the fact that the diffusion time is much shorter than the evolution times of interest allows us to consider the redshifted chemical imbalances, $\eta^\infty_\ell$, to be uniform throughout the NS core \citep{Reis1995}. These variables evolve as
\begin{eqnarray}
    \dot\eta_e^\infty=2W_{npe}\Omega\dot\Omega-Z_{npe}\Delta\tilde\Gamma^\infty_{npe}-Z_{np}\Delta\tilde\Gamma^\infty_{np\mu}, \label{eq:detae/dt}\\
    \dot\eta_\mu^\infty=2W_{np\mu}\Omega\dot\Omega-Z_{np}\Delta\tilde\Gamma^\infty_{npe}-Z_{np\mu}\Delta\tilde\Gamma^\infty_{np\mu},  \label{eq:detamu/dt}
\end{eqnarray}
where $\Omega=2\pi/P$ is the angular velocity of rotation of the NS, $\Delta\tilde\Gamma^\infty_{npe}$ and $\Delta\tilde\Gamma^\infty_{np\mu}$ are the total net rates at which neutrons are converted into protons and electrons and into protons and muons, respectively, and $W_{npe}$, $W_{np\mu}$, $Z_{npe}$, $Z_{np\mu}$, and $Z_{np}$ are constants that depend on the structure of the NS \citep{Fer2005,Reisenegger2006,Petro2010} and whose values for the models studied in the present paper are given in Table~\ref{table:parameters}.\footnote{Note that $W_{npe}<0$ and $W_{np\mu}<0$, therefore, when the star spins down ($\dot\Omega<0$), it makes the chemical imbalances increase. On the other hand, $Z_{npe}$, $Z_{np\mu}$, and $Z_{np}$ are all $>0$, and $\Delta\tilde\Gamma^\infty_{np\ell}$ has the same sign as $\eta_\ell$ for $\ell=e,\mu$, so the terms involving these quantities always make $|\eta_\ell|$ decrease.} The net effect of Urca reactions on the internal energy budget can be written as 
\begin{equation}
    L_{RH}^\infty-L_\nu^\infty=\sum_{\ell=e,\mu}h_\ell\eta_\ell^\infty\Delta\tilde\Gamma^\infty_{np\ell},
\end{equation}
where $h_\ell$ are dimensionless functions of the temperature, the chemical imbalances, and the NS structure and composition. For normal matter in the limit $|\eta_\ell|/k_B T_c\to\infty$, where $k_B$ is the Boltzmann constant, they satisfy $h_\ell\to 1/2$ for Durca reactions and $h_\ell\to 5/8$ for Murca reactions, representing the fraction of the chemical energy release that gets deposited as heat in the NS core, while the rest is emitted as neutrinos and antineutrinos, which escape. For $|\eta_\ell|/k_B T_c\lesssim 5$, one obtains $h_\ell<0$, corresponding to net cooling \citep{Fer2005}. 

Since the spin-down depends on the magnetic field, CPs increase their chemical imbalances much faster than MSPs \citep{GonRei2010}. For the strong magnetic field of CPs, the effect of rotochemical heating on temperature appears early (at $t\lesssim 10^5$ yr), while for the weaker magnetic field of MSPs it appears later ($t\gtrsim 10^{6-7}$ yr).

At later stages, the impact of compression due to spin-down on CPs becomes negligible, and their chemical imbalances only decrease slowly due to weak interactions, injecting heat and thus slowing down the cooling. MSPs, on the other hand, {might} reach a quasi-steady state where reactions counteract the rise in chemical imbalances caused by spin-down, 
\begin{equation}
    \Delta\tilde\Gamma^\infty_{np\ell,\mathrm{qs}}\approx-K_\ell\Omega\dot\Omega, \label{eq:qs rate}
\end{equation}
where 
\begin{eqnarray}
    K_e\equiv-\frac{2(Z_{np\mu}W_{npe}-Z_{np}W_{np\mu})}{Z_{npe}Z_{np\mu}-Z_{np}^2},\\
    K_\mu\equiv-\frac{2(Z_{npe}W_{np\mu}-Z_{np}W_{npe})}{Z_{npe}Z_{np\mu}-Z_{np}^2},
\end{eqnarray}
while heating and cooling effects also balance each other, yielding
\begin{equation}
    L_{\gamma,\mathrm{qs}}^\infty
    \approx -K\Omega\dot\Omega\bar\eta^\infty_{\mathrm{qs}}, 
    \label{eq:qs luminosity}
\end{equation}
where $K\equiv K_e+K_\mu$ 
and 
\begin{equation}\label{eq:qs eta}
    \bar\eta^\infty_{\mathrm{qs}}\equiv\frac{1}{K}
    \sum_{\ell=e,\mu}K_\ell h_\ell\eta^\infty_{\ell,\mathrm{qs}}
\end{equation}
\citep{Fer2005,Petro2010}. (Here and below, the subscript ``qs'' labels variables evaluated in the quasi-steady state.) Thus, the fraction of the spin-down power emitted as thermal photons is
\begin{equation}
    \frac{L_{\gamma,\mathrm{qs}}^\infty}{|\dot E_{\mathrm{rot}}|}=\frac{K\bar\eta^\infty_{\mathrm{qs}}}{I}\sim 3\times 10^{-4}\,\frac{\bar\eta^\infty_{\mathrm{qs}}}{\mathrm{MeV}},
    \label{eq:qs fraction}
\end{equation}
where $E_{\mathrm{rot}}=I\Omega^2/2$ is the kinetic energy of rotation and $I$ is the moment of inertia of the star. 

For a given NS structure with given weak interaction processes (Murca, Durca, or others, with or without Cooper pairing) and given NS spin-down parameters at a certain time, Eqs.~(\ref{eq:qs rate}) and (\ref{eq:qs luminosity}) determine the three quasi-steady-state parameters $\eta^\infty_{e,\mathrm{qs}}$, $\eta^\infty_{\mu,\mathrm{qs}}$ and $T^\infty_{s,\mathrm{qs}}$. In the relevant limit $k_BT^\infty_c\ll|\eta^\infty_\ell|$, the reaction rates become independent of $T^\infty_c$, so Eq. (\ref{eq:qs rate}) determines $\eta^\infty_{e,\mathrm{qs}}$ and $\eta^\infty_{\mu,\mathrm{qs}}$. Replacing these in Eq. (\ref{eq:qs luminosity}) yields the surface temperature $T^\infty_{s,\mathrm{qs}}$, independent of the NS envelope composition. 

Conversely, if a NS is in this quasi-steady state, the measured values of its spin-down parameters $\Omega$ and $\dot\Omega$ and its photon luminosity can be used in Eq.~(\ref{eq:qs luminosity}) to infer $\bar\eta^\infty_{\mathrm{qs}}$, without having to assume specific weak interaction processes. For the measured spin parameters and surface temperature of J0437 and the parameter $K$ of our Model M, one obtains $\bar\eta^\infty_{\mathrm{qs}}\approx 1.5\,\mathrm{MeV}$, but we note that $K$ varies by a factor $\sim 3$ between different equations of state, and the dependence on $(T_s^\infty)^4$ introduces an additional large uncertainty.

In order to reach the quasi-steady state from an initial chemical equilibrium state through the spin-down of the star (represented by the first term on the right-hand side of Eqs.~[\ref{eq:detae/dt}] and [\ref{eq:detamu/dt}]), the initial rotation rate $\Omega_0$ must satisfy
\begin{equation}\label{eq:reach qs}
    \Omega_0^2-\Omega^2\gtrsim\frac{K\bar\eta^\infty_{\mathrm{qs}}}{\sum_\ell(-W_{np\ell})K_\ell h_\ell}\approx\frac{L^\infty_{\gamma,\mathrm{qs}}}{\Omega\dot\Omega\sum_\ell W_{np\ell}K_\ell h_\ell}.
\end{equation}
Taking again the observed parameters of J0437 and the theoretical parameters from our Model M, and assuming $h_e\approx h_\mu\approx 1$, we obtain that the initial rotation period of this pulsar must satisfy $P_0\lesssim 1.6\,\mathrm{ms}$, substantially shorter than its present period $P=5.75\,\mathrm{ms}$, {but consistent with the range inferred from the age of its white dwarf companion (see Sect.~\ref{sec:observational}). In any case}, we reiterate that there are theoretical and observational uncertainties affecting these estimates.


When $|\eta_\ell^\infty|\gg k_BT_c^\infty$ and $|\eta_\ell^\infty|\gg\Delta_j^\infty$ for both $\ell=e,\mu$ and $j=n,p$ (which includes the case of normal matter out of chemical equilibrium, with $\Delta_j^\infty=0$), the net reaction rates become \citep{Reis1995,Petro2010}
\begin{eqnarray}
    \Delta\tilde\Gamma_{D,np\ell}^\infty\approx\frac{42}{457}\frac{\tilde L_{D,\ell}}{(\pi k_B)^6}(\eta_\ell^\infty)^5 \qquad \text{for Durca,} \label{eq:netDurca}\\
    \Delta\tilde\Gamma_{M,np\ell}^\infty\approx\frac{24}{11513}\frac{\tilde L_{M,\ell}}{(\pi k_B)^8}(\eta_\ell^\infty)^7 \qquad \text{for Murca,} \label{eq:netMurca}
\end{eqnarray}
where $\tilde L_{D,\ell}$, $\tilde L_{M,\ell}$, are the constants appearing in the neutrino luminosity for each process in chemical equilibrium, as given in Eq.~(\ref{eq:normalUrcaLum}). Setting the net rates from Eqs.~(\ref{eq:netDurca}) and (\ref{eq:netMurca}) equal to the quasi-steady-state rate for J0437 from Eq.~(\ref{eq:qs rate}) yields chemical imbalances substantially smaller than those required to account for the thermal photon luminosity of this pulsar. 


Thus, to account for the observed luminosity of J0437 by rotochemical heating, at least one of the nucleon species in the NS core must be Cooper-paired with a relatively large energy gap. \citet{Petro2010} included the effects of Cooper pairing with uniform energy gaps in the thermal and chemical evolution equations of \citet{Fer2005} (slightly corrected by \citealt{Reisenegger2006}) by introducing reduction factors, which are functions of the chemical imbalances and the temperature of the core. This work confirmed the arguments of \citet{Reis1997} in the sense that, in the Cooper-paired state, reactions are strongly suppressed as long as the chemical imbalances remain below the threshold energy,
\begin{equation}\label{thresh}
\Delta_{\text{thr}}=\begin{cases}
\Delta_n+\Delta_p & \text{for Durca reactions,} \\
\min\{\Delta_n + 3\Delta_p, 3 \Delta_n + \Delta_p\} & \text{for Murca reactions,} 
\end{cases}
\end{equation}
where $\Delta_n$ and $\Delta_p$ are the energy gaps of the neutrons and protons, allowing the chemical imbalances to grow freely up to this point. Once this threshold is reached, Urca reactions are activated, stopping further growth and rapidly heating the star, keeping it in a quasi-steady state, as in the case of normal matter. Therefore, the chemical imbalances remain near $\Delta_{\text{thr}}$, i.~e., at a value that can be much higher than in the normal case, releasing more energy in each reaction and thus leading to a higher quasi-steady temperature, given by Eqs.~(\ref{eq:qs luminosity}) and (\ref{eq:qs eta}) with $h_\ell\approx 1$ and $\eta^\infty_{\ell,\mathrm{qs}}\approx\Delta^\infty_{\mathrm{thr}}\approx\bar\eta^\infty_{\mathrm{qs}}$. This hypothesis can match the parameters of J0437 if $\Delta^\infty_{\mathrm{thr}}\approx 1.5\,\mathrm{MeV}$. (The value obtained by \citet{Petro2010} is lower, because they used a lower surface temperature for this pulsar.)

\citet{Gonj2015} extended the previous studies by considering the effect of density dependence and anisotropy of the energy gaps. They found that, for the case of the Murca process, the star reaches the quasi-steady state when the chemical imbalances exceed the smallest value of $\Delta_{\text{thr}}$ within the core, making the behaviour quite similar to the case with uniform gaps.

{On the other hand, \citet{Petro2011} found that, for Durca reactions with a large, uniform and isotropic energy gap, MSPs do not settle to a quasi-steady state, but their chemical imbalances and temperature oscillate strongly (on cooling time scales). Thus, in this case, it is not possible to make a reliable prediction about the temperature of these stars or to infer conclusions about the heating mechanisms from the observed temperature. We note, however, that the gaps are not expected to be uniform, but to vary with density. Therefore, the chemical imbalances will grow to the point where they reach approximately the minimum value of $\Delta_{\text{thr}}$, activating reactions in a thin shell where the gap takes this value, thus impeding further growth \citep{Gonj2015}. This reduces the effective volume of the core where reactions are active. We model this effect in Appendix \ref{appendix}, where we show that oscillations also happen for Murca reactions with a large enough, uniform energy gap. However, for both Durca and Murca, reducing the reaction rates by a small enough factor $f$, the oscillations are eliminated. In the simulations reported in the rest of this paper, we introduce such a factor to mimic the effect of a non-uniform gap and eliminate the oscillations.}

We note that none of this work considers lepton decay (conversion of electrons into muons, and vice-versa; \citealt{Alford2010}), a process that is intrinsically much slower than the Urca processes, but that is not suppressed by Cooper pairing, and which tends to reduce the difference between the electron and muon chemical potentials. Because Urca processes make these two chemical potentials evolve in a very similar way, their differences generally remain small anyway. Thus, this process is not expected to substantially change the results \citep{Kantor2021} and will also be ignored in the present work.


Another effect not considered in the work described above, but initially pointed out by \citet{Schaab1998} and recently discussed and evaluated by \citet{Kantor2021}, is that, when the protons undergo the transition to the superconducting state, the magnetic flux gets concentrated into small regions in which the protons continue behaving as normal (non-superconducting) particles. Since this introduces an additional parameter, namely the fraction of the total volume in these regions, we do not consider it here (effectively assuming perfect flux expulsion from the superconductor), but we expect to address it in a forthcoming paper.

\subsection{Vortex Creep}

In the inner crust, the superfluid neutrons form an array of discrete quantum vortices, which are pinned to the nuclei of the solid nuclear lattice, generally keeping the superfluid rotating faster than the rest of the star. As the star spins down, the relative velocity between the superfluid and the vortex array
grows, inducing a Magnus force. When a critical velocity difference is reached between the superfluid and the vortex array, the Magnus force overcomes the pinning force, and the pinning and unpinning of many vortices releases energy that heats the star, with a total time-averaged power \citep{Alpar1984} 
\begin{equation}
L_{\text{vc}} = J |\dot\Omega|, 
\label{eq:vortex}
\end{equation}
where $J$ is the excess angular momentum, given by $J\approx I_p\widetilde{\omega}_{cr}$, $I_p$ is the moment of inertia of the superfluid in the pinning layer, and $\widetilde{\omega}_{cr}= (\Omega_s-\Omega_c)_{cr}$ is the average over this layer of the critical lag between the angular velocity of the crust, $\Omega_c$, and the macroscopic angular velocity of the superfluid, $\Omega_s$, and $\dot\Omega$ is the time-derivative of the star's angular velocity (almost the same for the crust and the superfluid). The value of $J$ is expected to be in the range $\sim 10^{43-45}$ erg s, so 
\begin{equation}
L_{\text{vc}} \sim (10^{29}-10^{31})\, |\dot{\Omega}_{-14}|\, \text{erg s}^{-1},
\label{eq:vortexL}
\end{equation}
with $\dot{\Omega}_{-14}=\dot\Omega/(10^{-14}$ s$^{-2})$ \citep{GonRei2010}. If, at a late time $t\gtrsim 10^6\,\mathrm{yr}$ in the evolution of a NS, vortex creep is the dominant heating mechanism, its photon luminosity will be given by Eqs.~(\ref{eq:vortex}) and (\ref{eq:vortexL}). {We note that this was used by \citet{Fujiwara2024VC} to obtain values of $J$ for many pulsars. However, some of those values differ by orders of magnitude, which we do not expect from NS models, making it likely that different heating mechanisms are at work.}

\subsection{Crustal heating}

When NSs accrete matter from a binary companion, their crust gets compressed, inducing some nuclear transformations and leaving some layers of matter close to the threshold for such transformations to occur. After accretion has stopped, the decrease of the centrifugal force due to the continuous spin-down of the star causes an additional small compression that can make the matter cross this threshold, inducing further exothermic reactions (electron captures, neutron emissions, or pycnonuclear reactions) at the interfaces between the deep layers of the crust with different dominant nuclei, releasing heat at a rate \citep{Gus2015} 
\begin{equation}
   L_{\text{CH}} \approx 7\times 10^{28}a\frac{R_6^7}{M_{1.4}^2}\Omega_3 |\dot\Omega_{-14}| \sum_i \frac{p_i}{10^{31}\text{erg cm}^{-3}} \frac{q_i}{1\text{MeV}} \,\text{erg s}^{-1}.
	\label{eq:ch}
\end{equation}
Here $R_6= R/(10^6\mathrm{cm})$,  $R$  is the stellar radius, $M_{1.4} = M/(1.4 M_{\odot})$, $M$ is the stellar mass, $\Omega_3=\Omega/(10^3 \mathrm{s}^{-1})$, $\Omega$ is the angular velocity, and $a$ is a positive, dimensionless parameter of order unity. The various transitions between nuclear species in the crust of the NS, labeled by index $i$, occur at critical pressures $p_i$ and can release an energy $q_i$ per nucleon, whose respective values are listed in \citet{Hae90}. 
We consider all reactions to determine the upper limit of temperatures that this mechanism can predict, resulting in the following expression:
\begin{equation}
   L_{\text{CH}} \approx 1.32\times 10^{30}\Omega_3 |\dot\Omega_{-14}|  \,\text{erg s}^{-1}.
	\label{eq:ch}
\end{equation}

As mentioned in Sect.~\ref{sec:intro}, \citet{Gusakov2020a} suggested that this effect might be much weaker. Therefore, we regard Eq.~(\ref{eq:ch}) as an upper limit. We consider this mechanism only for MSPs, which are generally accepted to have been recycled by accretion.

\section{Thermal evolution simulations for the observed NSs} \label{sec:simulations}

\subsection{Assumptions and setup}\label{sec:setup}

To identify and test the dominant heating mechanisms, we compared the predictions of thermal evolution models to the HST observations. The thermal evolution curves were calculated by integrating Eqs.~(\ref{ThermalEvol}), (\ref{eq:detae/dt}), and (\ref{eq:detamu/dt}) with a modified version of the code of \citet{Petro2010}, under the assumption of magnetic dipole spin-down with a constant magnetic field strength $B=3.2\times 10^{19}(P[\mathrm{s}]\dot P)^{1/2}$ G inferred from the observed values of the period $P$ and its time-derivative $\dot P$ for the respective pulsar. The initial period was set to $P_0=1$ ms for MSPs and $P_0=5$ ms for CPs. 
For CPs we assumed a fully catalyzed envelope containing heavy elements \citep{Gud83}, whereas for MSPs we assumed an accreted envelope with light elements \citep{Potekhin1997}, but we remind the reader that the late-time evolution of $T_s^\infty$ does not depend on this choice.

For the CPs (B0950, J0108, and J2144), we assumed an initial core temperature $T_0^{\infty}=10^{11}$ K for all curves. For the MSPs (J0437 and J2124) we set 
$T_0^{\infty}=10^{9}$ K, 
roughly matching the conditions expected at the end of the accretion phase. We verified that, for both cases, any choice of $T_0\gtrsim 10^9\,\mathrm{K}$ does not affect our results, since the initial neutrino emission is so strong that it causes the temperature to drop to $10^9$ K in less than one year. When considering rotochemical heating, we set the initial values for the chemical imbalances equal to zero, which is irrelevant to our results as, for the magnetic field strengths considered here, evolution curves for different initial values rapidly converge
\citep{Reis1995,Fer2005,Kantor2021}. 

\begin{figure*}
\centering
    \includegraphics[width=0.5\linewidth]{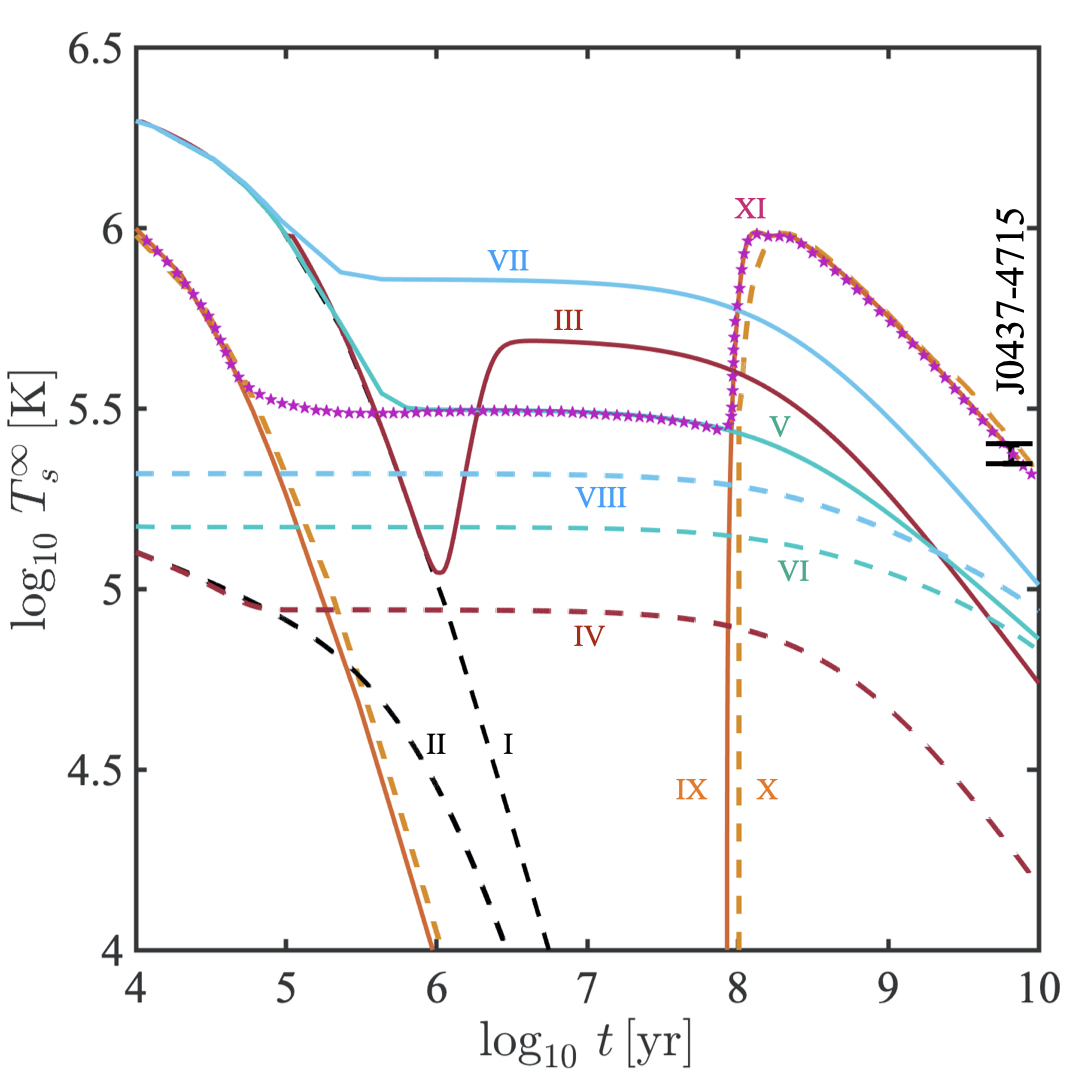}\hfil
    \includegraphics[width=0.5\linewidth]{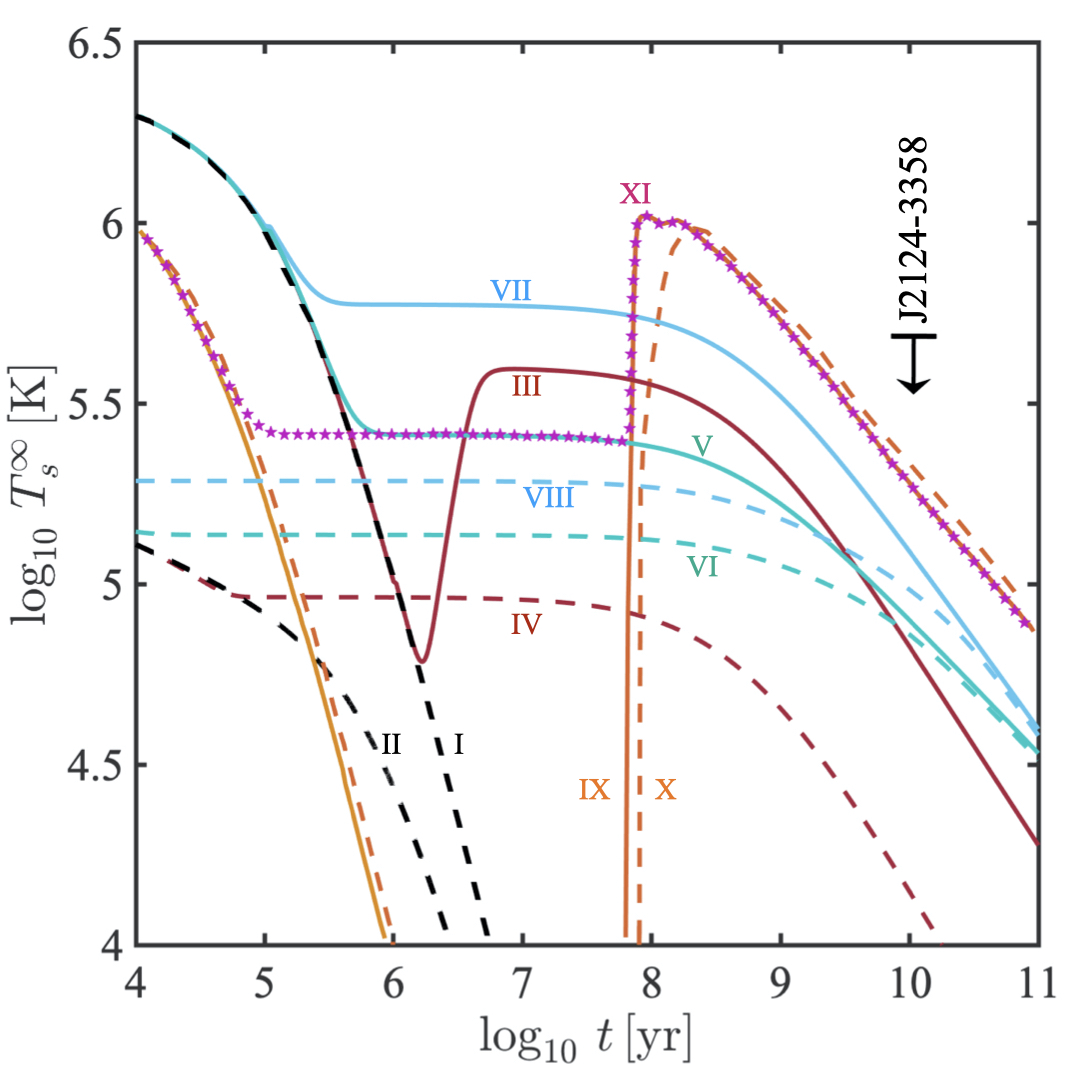}\par\medskip
    \includegraphics[width=0.5\linewidth]{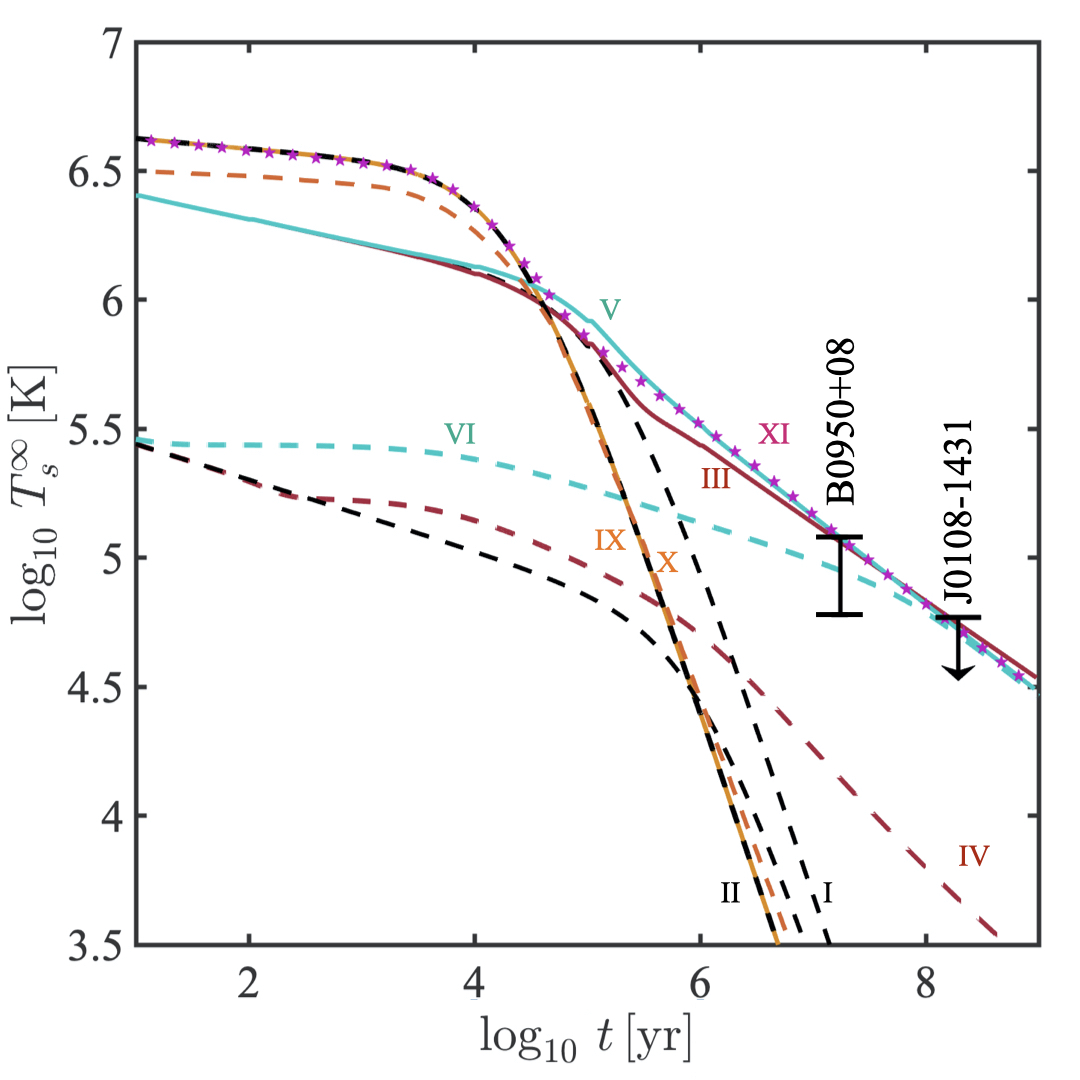}\hfil
    \includegraphics[width=0.5\linewidth]{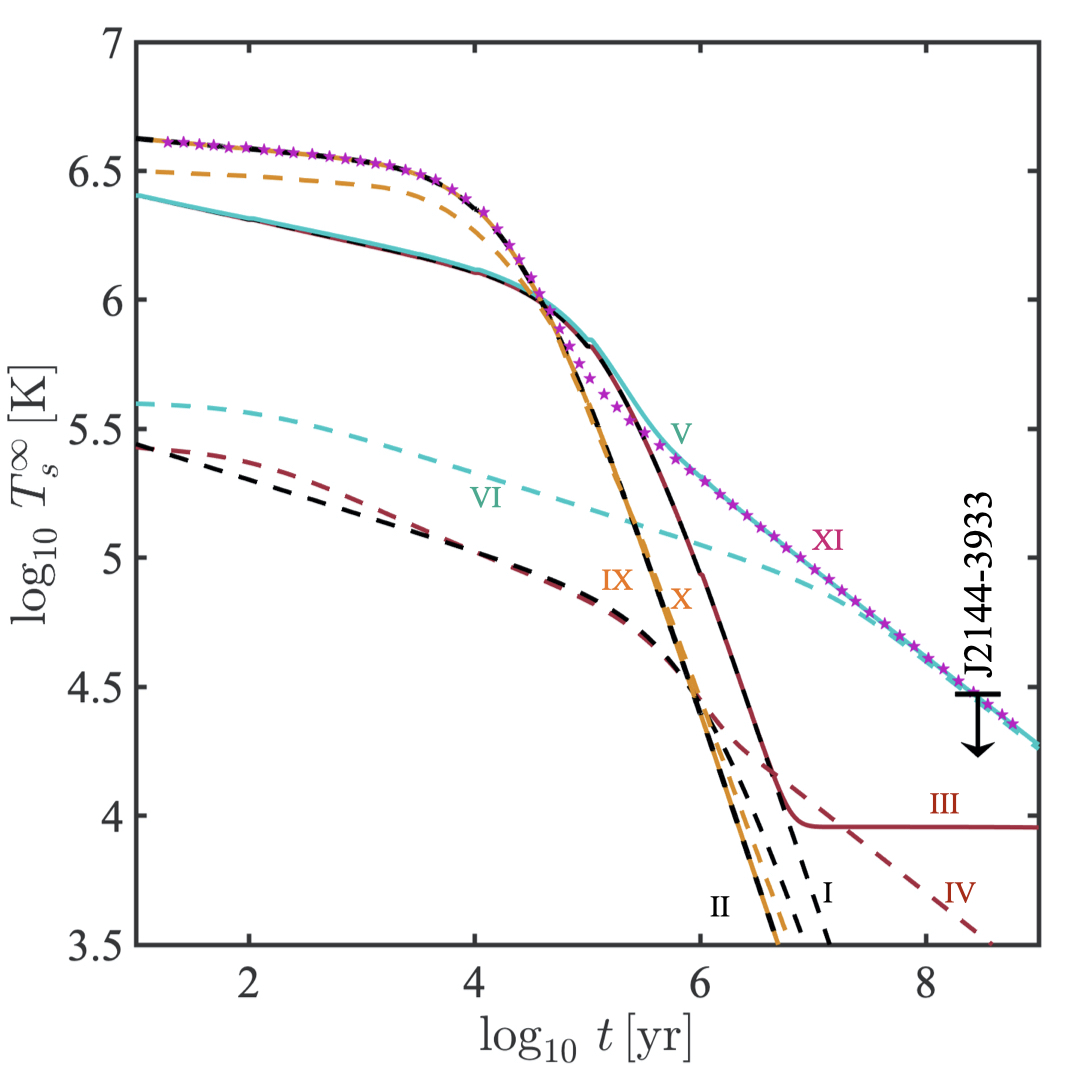}\hfil
\caption{Surface temperature evolution curves for each observed pulsar considering models with different reheating mechanisms. In each panel, all curves were calculated assuming magnetic dipole spin-down with a constant dipole moment inferred from the measured spin-down parameters $P$ and $\dot P$ of the respective pulsars. B0950 and J0108 are plotted together, because they have nearly the same inferred dipole field. Their curves were generated with the magnetic field of B0950. For the MSPs (J0437 and J2124), the initial conditions were taken as $P_0=1$ ms and $T_0^\infty=10^{9}$ K, whereas for the CPs (B0950, J0108, and J2144) we assumed $P_0=5~\mathrm{ms}$ and $T_0^\infty=10^{11}$ K. The error bars or upper limits for the pulsars are placed at the times at which the present spin parameters are reached. The models used for Murca and Durca reactions are given in Table~\ref{table:parameters} as Model M and Model D, respectively.
The labels correspond to the following cases: I- Passive cooling with Murca reactions in normal (non-superfluid and non-superconducting) matter. II- Passive cooling with Durca reactions in normal matter.  III- Rotochemical heating with Murca reactions in normal matter. IV- Rotochemical heating with Durca reactions in normal matter. V- Vortex creep with Murca reactions setting $J=3\times 10^{43}$ erg s and assuming normal particles in the core. VI- Same as V, but with Durca reactions.  VII- Crustal heating (only present in MSPs) with Murca reactions and normal particles in the core. VIII- Same as VII, but with Durca reactions.  IX- Rotochemical heating with Murca reactions considering normal protons and superfluid neutrons with a uniform energy gap $\Delta_n=1.5$ MeV, {with reaction rates reduced by a factor $f=10^{-2}$.} X- Same as IX, but with Durca reactions {and reaction rates reduced by a factor $f=10^{-8}$.} XI- Rotochemical heating with Murca reactions considering normal protons and superfluid neutrons with $\Delta_n=1.5\,\mathrm{MeV}$ {and reaction rates reduced by a factor $f=10^{-2}$}, as well as vortex creep with $J=3\times 10^{43}$ erg s.
}\label{graficos}
\end{figure*}

\begin{table*}
\caption{Cases simulated.} 
\begin{center}
\begin{tabular}[H]{ccccccc}
\hline

Case & 
Murca/Durca   & 
Normal/Superfluid   & 
Passive & 
Rotochemical &
Vortex Creep &
Crustal 
\\
\hline
I & M & N & X & & &  \\ 
II & D & N & X & & & \\ 
\hline
III & M & N & & X & & \\ 
IV & D & N & & X & & \\ 
\hline
V & M & N & & & X &  \\ 
VI & D & N & & & X &  \\ 
\hline
VII & M & N & & & & X  \\ 
VIII & D & N & & & & X  \\ 
\hline
IX & M & S & & X & &  \\ 
X & D & S & & X & &  \\ 
\hline
XI & M & S & & X & X &  \\ 

\hline

\hline
\end{tabular}
\tablefoot{The first column corresponds to the labelling of the curves in Fig.~\ref{graficos}. The second indicates whether Durca (D) or only Murca (M) reactions are allowed. The third distinguishes between normal (N) and superfluid (S) neutrons. (The protons are always taken to be normal.) An X in the fourth indicates passive cooling (no heating mechanisms). An X in any of the last three columns indicates that the respective heating mechanism is present.}   \label{table:sims}
\end{center}
\end{table*}

\subsection{Results of simulations and comparison to observations}\label{sec:results}

In Fig.~\ref{graficos}, we show thermal evolution simulations for each of the observed pulsars, considering different internal heating mechanisms (or none) and the normal or Cooper-paired states of the nucleons in the core, as summarized in Table~\ref{table:sims}. The CPs B0950 and J0108 were combined in the same panel, because they have very similar magnetic fields, therefore, as long as they had similar initial rotation periods and internal properties, they could be regarded as two different points along the same thermal evolution curves, for whose calculation we considered the magnetic field value inferred for B0950. We plotted the data point for the measured temperature or upper limit of each pulsar at the time $t$ at which the computed spin-down parameters $P(t)$ and $\dot P(t)$ match the observed values. This time is in general somewhat shorter than the characteristic age, $\tau\equiv P/(2\dot P)$, which would agree in the limit of infinitely rapid initial rotation. 

One can see that the curves without reheating (labeled I and II) drop to very low temperatures, $T_s^\infty\sim 10^4\,\mathrm{K}$ at early times $t<10^7\,\mathrm{yr}$, so they cannot match the relatively high temperatures $T_s^\infty\sim 10^5\,\mathrm{K}$, of the observed pulsars. This is shown here for non-Cooper-paired matter, but remains true for any plausible assumptions regarding Cooper pairing of neutrons and/or protons \citep{Page2013}. Thus, some heating mechanism is required to explain the observations. 

For rotochemical heating in CPs containing normal matter, which is discussed and illustrated in some detail in \citet{GonRei2010} and labeled III and IV in the lower panels of Fig.~\ref{graficos}, neutrino emission is the primary cooling mechanism from the NS's birth until $t\sim 10^4-10^5$ years. During this period, the chemical imbalances (shown in Fig.~3 of \citealt{GonRei2010}) increase as a result of the star's spin-down, reaching a maximum that largely depends on the initial period and the strength of the external dipole field. Beyond the neutrino cooling age, the NS cools down through thermal photon emission from its surface. At this stage, the effect of spin-down becomes negligible, but the slow release of the previously accumulated chemical imbalances makes the temperature decrease much more slowly than in the absence of heating. 
We have verified that curves with different initial periods asymptotically converge, because the late evolution of the chemical imbalances in CPs is fully determined by the beta reactions, while the temperature is determined by the balance of the heat input through these reactions against the cooling through photons from the NS surface (Eq.~\ref{heating=cooling}).

For rotochemical heating in MSPs with normal matter, previously studied by \citet{Reis1995}, \citet{Fer2005}, and \citet{GonRei2010} and labeled III and IV in the upper panels of Fig.~\ref{graficos}, the thermal evolution is slightly different. From the end of recycling ($t=0$) until $t\sim 10^4-10^5\,\mathrm{yr}$, neutrino emission is the dominant cooling mechanism, and the chemical imbalances increase due to spin-down. However, the latter happens more slowly than in CPs, because of the weaker dipole field, so the imbalances do not saturate during this period. As the star transitions to the photon cooling age, the chemical imbalances continue to grow until reaching a quasi-steady state in $t\sim 10^6-10^7$ yr. At this stage, the spin-down remains important, but it is balanced by the effect of the reactions (Eq.~\ref{eq:qs rate}), so the NS reaches the quasi-steady state discussed in Sect.~\ref{sec:rotochemical}, and the chemical imbalances remain roughly constant. Similarly, the heat input from the reactions is again balanced by the cooling effect through thermal photon emission from the NS surface (Eq.~\ref{heating=cooling}).

Rotochemical heating with Murca reactions in normal matter (III) can account for the observed temperatures of the CPs B0950 and J0108, and is consistent with the upper limits for the CP J2144 and the MSP J2124. However, as argued in Sect.~\ref{sec:rotochemical}, it is insufficient to explain the high temperature of MSP J0437. Rotochemical heating with Durca reactions (IV), whose physics in principle is quite similar to the Murca scenario, produces faster beta reactions in normal matter, which 
leads to the saturation of the chemical imbalances at lower values and, consequently, cooler temperatures compared to Murca reactions. Any choice of initial period, initial chemical imbalances, and NS model for Durca reactions in normal matter is found to underestimate the observed temperatures.

For rotochemical heating in a NS core with superfluid neutrons, superconducting protons (with no magnetic flux penetrating them), or both, illustrated by curves IX and X in all panels, beta reactions are suppressed as long as the chemical imbalances remain below the threshold $\Delta_{\mathrm{thr}}$ given by Eq.~(\ref{thresh}). As explained in Sect.~\ref{sec:rotochemical}, this threshold must be rather high, $\Delta_{\mathrm{thr}}\sim 1.5\,\mathrm{MeV}$, to explain the thermal photon emission of J0437. For the imbalances to reach this high threshold, the initial rotation of the pulsar must be fast, $P_0\lesssim 2\,\mathrm{ms}$, which is plausible for MSPs, but not for CPs. Since, in our simulations, we chose $P_0=1\,\mathrm{ms}$ for the MSPs, $P_0=5\,\mathrm{ms}$ for the CPs, and $\Delta_{\mathrm{thr}}=1.5\,\mathrm{MeV}$ for all, we find that the MSPs reach the threshold, so rotochemical heating is activated, explaining the thermal emission of J0437, whereas CPs do not reach the threshold and therefore are not reheated, cooling down to low temperatures. 
Thus, it is not possible to explain all the observations with rotochemical heating as the only heating mechanism, unless a small amount of magnetic flux is present inside a core with superconducting protons, which we expect to study in a forthcoming paper. We note that in the simulations with Cooper pairing we assumed uniform and isotropic gaps, but reduced the reaction rates by constant factors $f=10^{-2}$ for Murca and $f=10^{-8}$ for Durca reactions, which are roughly the maximum values that suppress the oscillations studied in \citet{Petro2011} and in Appendix~\ref{appendix}. Any smaller values do not affect the results significantly. These reductions somehow mimic the effect of non-uniform energy gaps, which would only allow reactions in some thin shell within the NS core where $\Delta_{\mathrm{thr}}$ is smallest \citep{Gonj2015}. 

Another potentially important heating mechanism is vortex creep, represented as curves V and VI in all panels of Fig.~\ref{graficos}. This mechanism leads to significant deviations from standard cooling models once a star reaches the photon cooling age in CPs and somewhat later in MSPs. The heat generated by vortex friction in the crust is proportional to the star's angular deceleration $|\dot{\Omega}|$, which is generally larger in CPs than in MSPs. 
In our analysis of vortex creep, we treat the uncertain excess angular momentum, $J$, as a free parameter, which is constrained by the upper limit for J2144, corresponding to $J\leq 3\times 10^{43}$ erg s. Adopting this upper bound for both of our vortex creep models (V and VI), it can account for the temperature of B0950 but is insufficient for J0437. Thus, if $J$ is approximately universal for all NSs (as we expect since it is determined by the structure of the NS crust), vortex creep cannot by itself account for all the observations. 

Crustal heating (labeled VII and VIII in the upper panels of Fig.~\ref{graficos}) occurs only in recycled pulsars, that is, MSPs. Similarly to vortex creep, the quasi-steady temperature in this case does not depend on the type of Urca reactions that occur in the core, but is solely determined by the pycnonuclear reactions in the deeper layers of the crust and the current spin-down of the NS, quantified by $\Omega \dot{\Omega}$. We included all possible pycnonuclear reactions listed in \cite{Hae90} to establish an upper limit for this mechanism. Both MSPs achieve very similar temperatures because they have comparable spin-down rates. However, this upper limit is too low to account for the observed temperature of J0437.

In both vortex creep and crustal heating, the late-time temperature is set by a balance between the respective heating effect and surface photon emission. Thus, it depends only on the current spin parameters of the pulsar and is unaffected by the type of Urca reactions and the Cooper pairing in the core. 

Since rotochemical heating with Cooper-paired nucleons with a large gap can account for the surface temperature of J0437 and vortex creep for that of B0950, but not vice-versa, we also model the combination of both effects (each with the same parameters as before), which is labeled XI in all panels. It can be seen that, at any given time, the temperature is roughly the higher one among the two obtained by considering each effect independently. At late times, rotochemical heating dominates for MSPs, explaining the temperature of J0437, and vortex creep dominates for CPs, explaining that of B0950. 
It is interesting to note that this model also predicts that the temperatures of the other pulsars should be close to the upper limits inferred from their HST observations. Thus, the emission detected for J2124 and possibly detected for J0108 should have an important thermal contribution, and the thermal emission of J2144 should not be much lower than the upper limit from its non-detection.

\subsection{Sensitivity to initial conditions}

As mentioned in Sect.~\ref{sec:thermal_evolution}, NSs are born with very high core temperatures, $T_c\sim 10^{11}\,\mathrm{K}$. Although not quite as extreme, MSPs also emerge from their accretion phase with a high core temperature, $T_c\sim 10^{8-9}\,\mathrm{K}$. In the early neutrino cooling stage, for $T_c\gtrsim 10^8\,\mathrm{K}$, the luminosity is a strong function of temperature, $\propto T_c^6$ for Durca and $\propto T_c^8$ for Murca reactions (assuming normal matter), so the NS quickly loses any memory of its initial temperature. The pulsars included in the present study have surface temperatures $T_s\lesssim 2\times 10^5\,\mathrm{K}$, corresponding to core temperatures of at most a few times $10^6\,\mathrm{K}$, so their present state should be independent of their initial temperature. 

Similarly, it has been pointed out that, at such high initial temperatures, any initial chemical imbalance is quickly erased \citep{Reis1995,Fer2005}, justifying our assumption of vanishing initial values for $\eta_e$ and $\eta_\mu$, which we also verified through simulations with non-zero initial values.

The effect of the initial rotation period $P_0$ is less obvious. First, we note that 
\begin{equation}
    \frac{P_0^2}{P^2}=1-\frac{t}{\tau}.
\end{equation}
Thus, as long as $t\lesssim\tau$, we have $P\sim P_0$. During this time, the rotational properties of the pulsar do not vary much, so the probability of detecting it as a radio pulsar is also roughly constant, and the probability to detect it at an age $<t$ is $\sim t/\tau$. For the characteristic ages $\tau\sim 10^{7-10}\,\mathrm{yr}$, it is very unlikely for any of them to be young enough to explain its temperature through passive cooling, which requires $t\lesssim 10^5\,\mathrm{yr}$. 

If vortex friction or crustal heating dominates, the luminosity at $t\gg 10^5\,\mathrm{yr}$ depends just on the present spin-down parameters, $\dot\Omega$ or $\Omega\dot\Omega$, respectively, not on the initial conditions. This imposes a very weak constraint on the initial period, which only needs to be slightly shorter than the present one. 

For rotochemical heating with Cooper-paired nucleons, the initial period can have a strong qualitative effect by determining whether the chemical imbalances can grow enough to reach the threshold $\Delta_{\mathrm{thr}}$ and thus activating the internal heating. For this to happen, it requires
\begin{equation}
    \frac{1}{P_0^2}-\frac{1}{P^2}\gtrsim\frac{1}{(1.8\,\mathrm{ms})^2}\frac{\Delta_{\mathrm{thr}}}{1.5\,\mathrm{MeV}}.
\end{equation}
For MSPs, it appears possible to satisfy this condition, as there are a few known MSPs with periods $P\approx 1.5\,\mathrm{ms}$ (e.~g., \citealt{Backer1982,Hessels2006,Bassa2017,Ho2019}), although it is not obvious that all MSPs have had such fast rotations. Among CPs, the fastest known has a period $P=16\,\mathrm{ms}$ \citep{Marshall1998}, therefore it is very unlikely for any of them to satisfy this condition. 
For J0437, as discussed in Sect.~\ref{sec:observational}, its age inferred from the temperature of its white dwarf companion, $t=6.0\pm 0.5\,\mathrm{Gyr}$ \citep{Durant2012}, is quite close to its (intrinsic) characteristic age, $\tau=6.7\pm 0.2\,\mathrm{Gyr}$, suggesting that its ``initial'' period (as it emerged from the accretion phase and started spinning down), $P_0$, was substantially shorter than its current period $P$. {This makes it plausible, but by no means certain, that $P_0$ was short enough for the chemical imbalances to reach $\Delta_{thr}$, activating rotochemical heating and explaining the observed temperature. For J2124, the absence of a binary companion does not allow such an assessment.} 



\section{Conclusions}\label{sec:conclusions}
 
Passive cooling models of NSs predict surface temperatures $T_s^\infty\ll 10^5\,\mathrm{K}$ at ages $t>10^6\,\mathrm{yr}$. In the present paper, we have considered deep UV observations of five pulsars with characteristic ages much larger than this, four of which have detectable emission. In two cases (MSP J0437 and CP B0950), this can be confidently interpreted as thermal emission from their whole surface, leading to measurements of $T_s^\infty\sim 10^5\,\mathrm{K}$, which imply the presence of some heating mechanism. For the other three cases (MSP J2124 and CPs J0108 and J2144), we considered upper limits on $T_s^\infty$ based on their UV detection or non-detection.

We modeled the thermal evolution of NSs with three potentially important heating mechanisms, namely rotochemical heating, vortex creep, and crustal heating, with and without Cooper pairing of neutrons and/or protons, but ignoring the possible presence of magnetized structures within the superconducting protons. We found that none of the heating mechanisms by itself can match all the observations. However, it is possible to account for all observational data with a model that combines vortex creep with rotochemical heating in the presence of a large, {uniform\footnote{{We remind that, as found in Appendix~\ref{appendix}, MSPs with rotochemical heating in the presence of a large, uniform energy gap will enter a cycle of oscillations of their chemical imbalances and temperature, making their surface temperature essentially unpredictable. However, the energy gaps are not expected to be uniform, with the consequence that reactions should only be activated in a relatively thin shell. We modeled this by introducing a small factor $f$ multiplying the reaction rates, which eliminates the oscillations.}} and isotropic} Cooper pairing gap $\approx 1.5\,\mathrm{MeV}$ in one of the nucleon species (or {in both of them, with} an equivalent combination $\Delta_{\mathrm{thr}}\approx 1.5\,\mathrm{MeV}$) and a fast initial rotation ($P_0\lesssim 2\,\mathrm{ms}$) for J0437. In this model, rotochemical heating is the dominant heating mechanism in MSPs, getting activated as they spin down enough for their chemical imbalances to exceed $\Delta_{\mathrm{thr}}$, thus explaining the temperature of J0437. CPs, on the other hand, are unlikely to have a fast enough initial rotation for this to happen, but the temperature of B0950 can be explained by vortex creep. In this model, the predicted temperatures for the three other pulsars are close to their observational upper limits, making further observations in multiple filters worthwhile.

\begin{acknowledgements}
This work was supported by ANID-FONDECYT grant 1201582 and the Center for Astrophysics and Associated Technologies (CATA; ANID Basal grant FB210003). 
\end{acknowledgements}

\bibliographystyle{aa}
\bibliography{biblio}


\begin{appendix}

\section{Thermal oscillations}\label{appendix}

\begin{figure*}
\centering
    \includegraphics[width=0.5\linewidth]{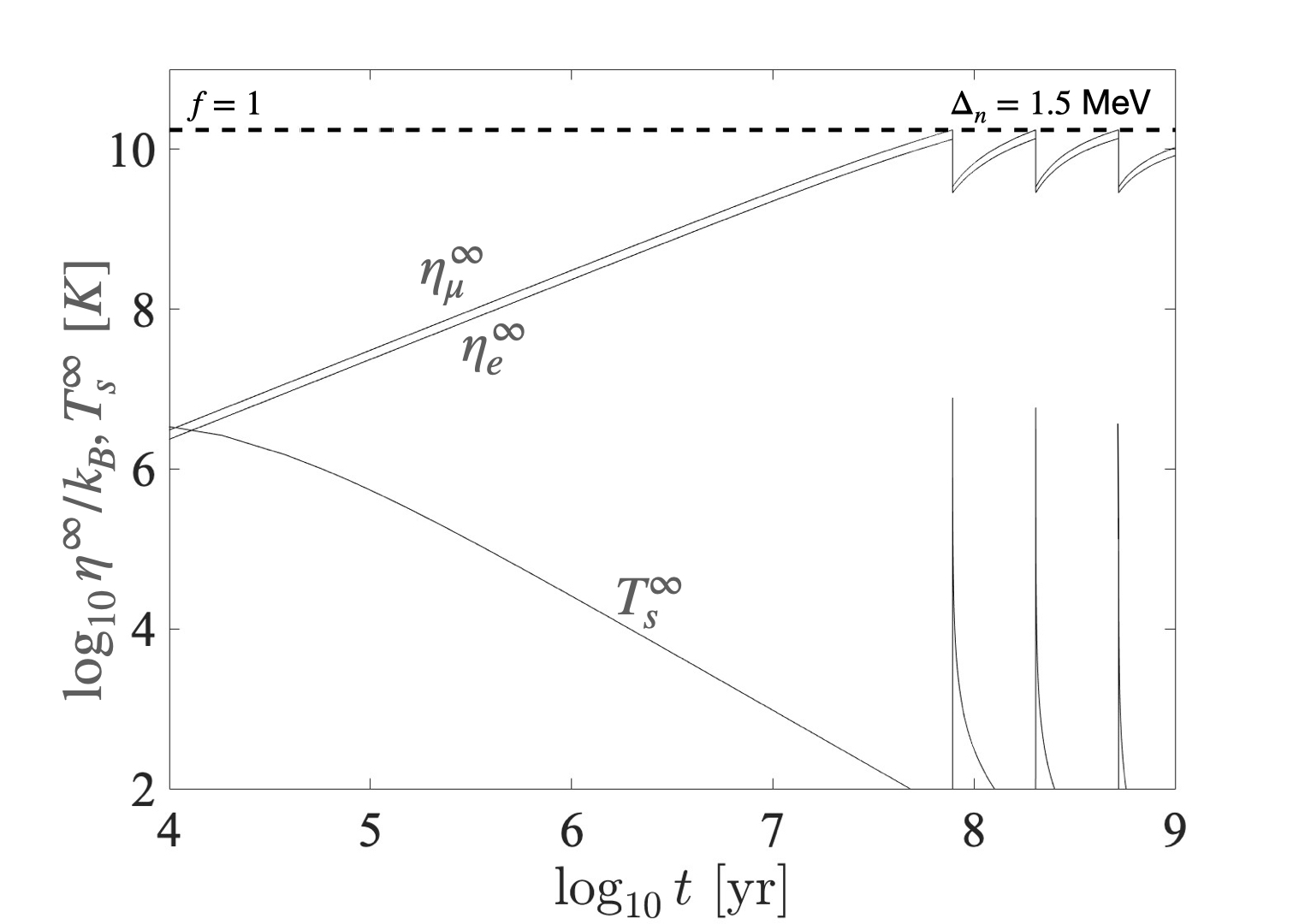}\hfil
    \includegraphics[width=0.5\linewidth]{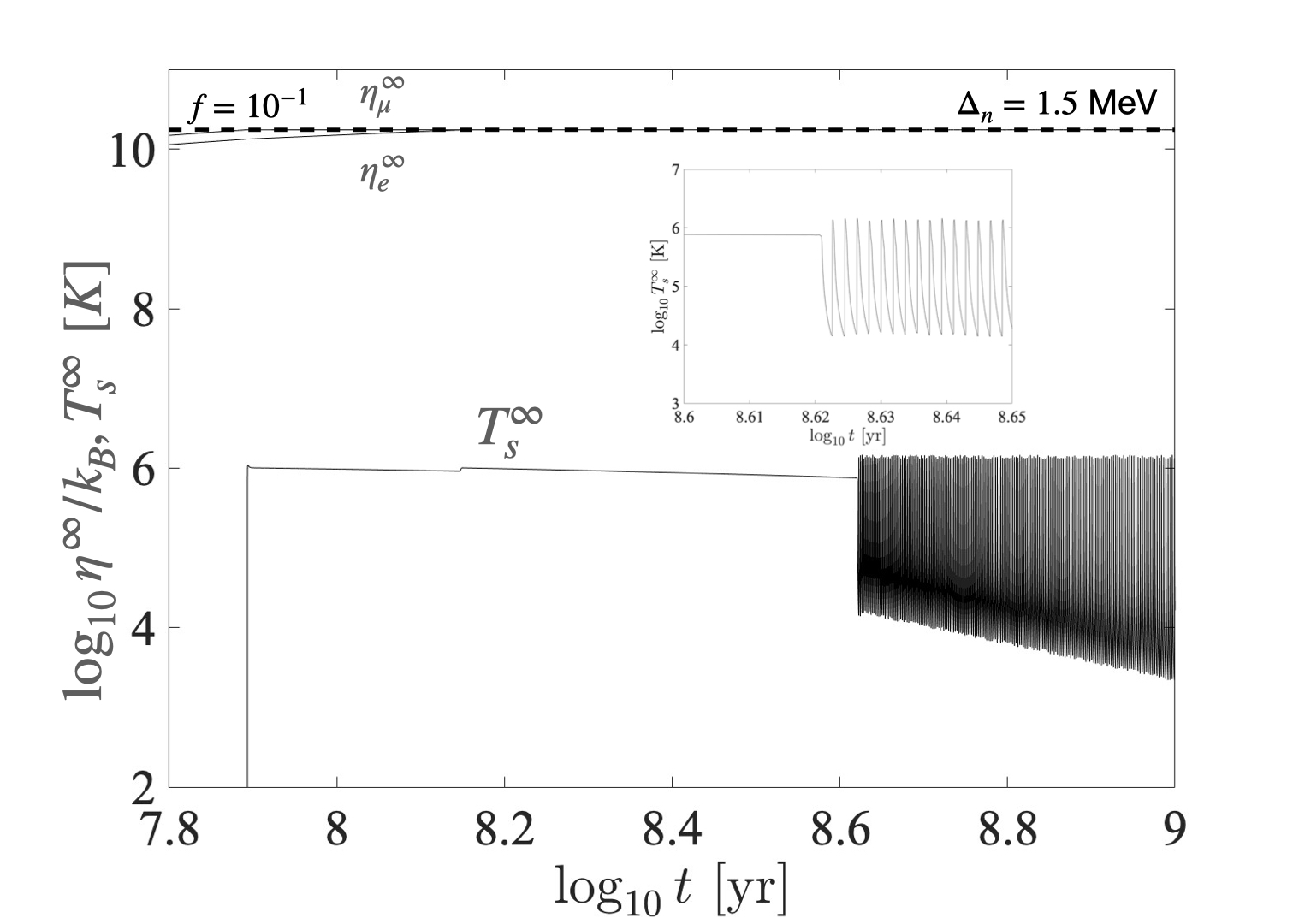}\par\medskip
    \includegraphics[width=0.5\linewidth]{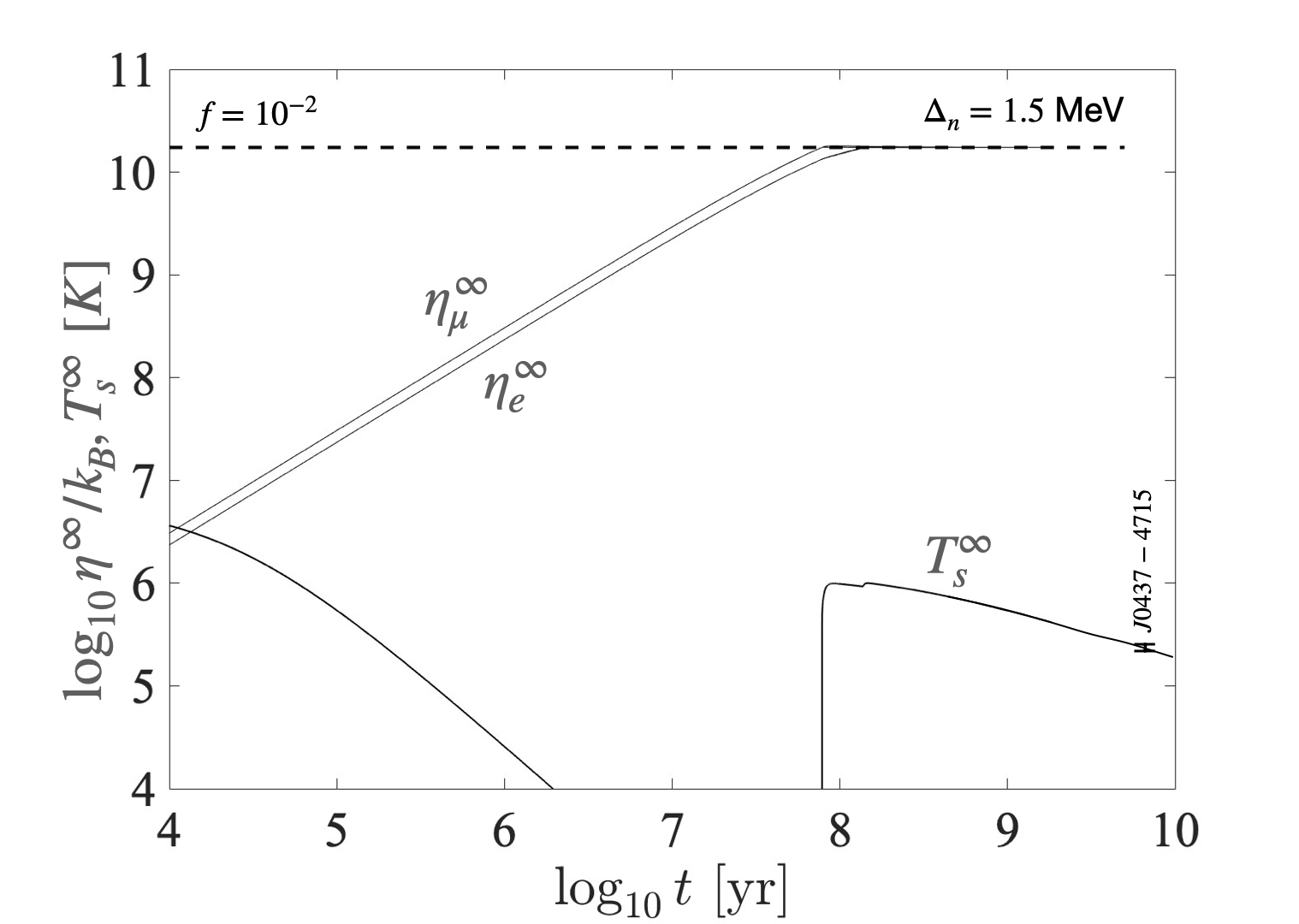}\hfil
    \includegraphics[width=0.5\linewidth]{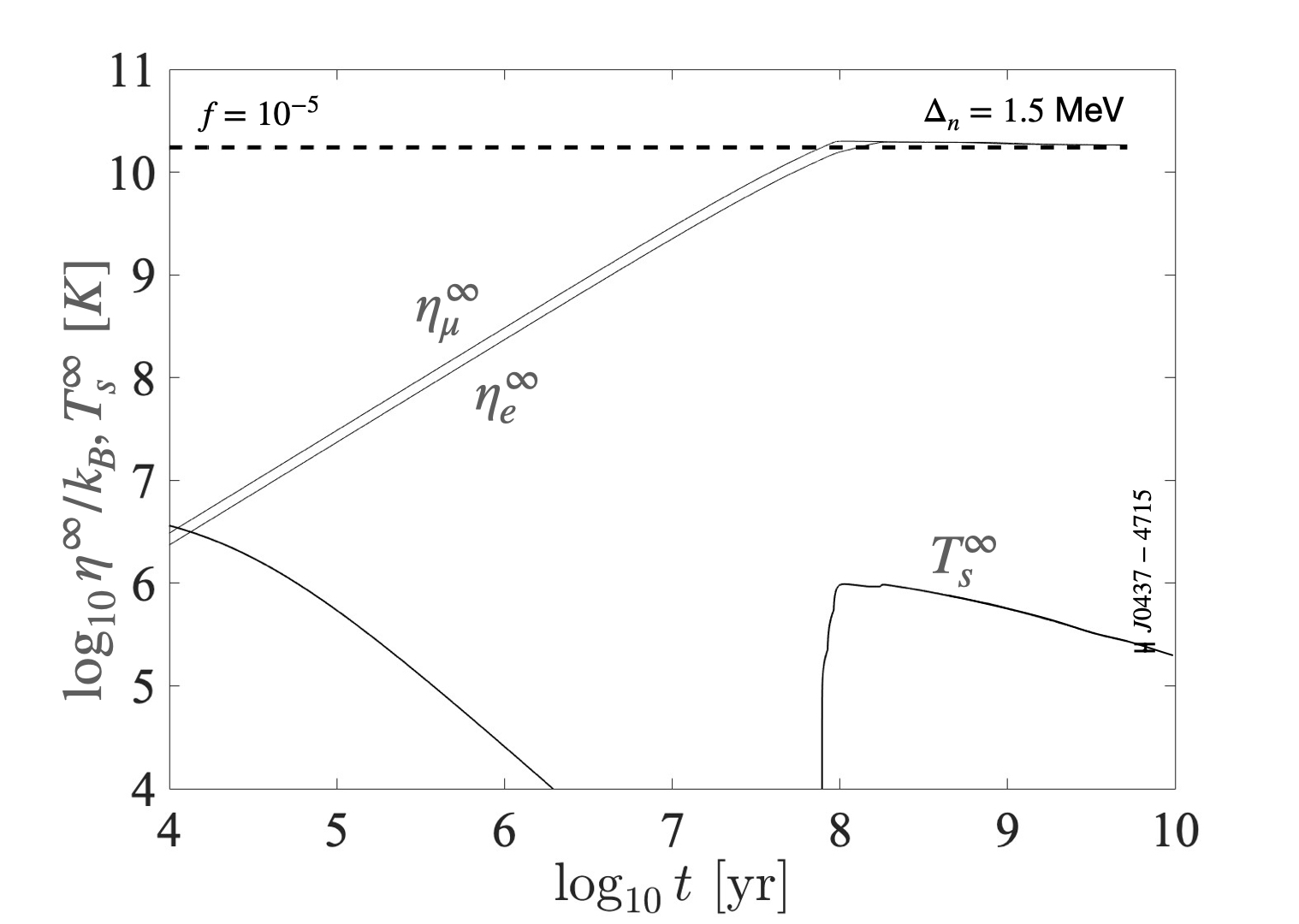}\hfil
\caption{{Evolution of the chemical imbalances and the temperature for an MSP with initial conditions $P_0=1\,\mathrm{ms}$, $T_0^\infty=10^9\,\mathrm{K}$, and $\eta_{e,0}^\infty=\eta_{\mu,0}^\infty=0$, considering the effect of rotochemical heating with Murca reactions, normal protons, and superfluid neutrons with $\Delta_n^\infty=1.5\,\mathrm{MeV}$. Each panel has the reaction rates reduced by a different factor $f=1,10^{-1},10^{-2},10^{-5}$, respectively.}
}\label{oscillations}
\end{figure*}

As mentioned in Sect.~\ref{sec:rotochemical}, {\citet{Petro2011} found that the evolution of an MSP with rotochemical heating in the presence of Durca reactions and a large, uniform, and isotropic Cooper pairing gap leads to strong oscillations of the chemical imbalances and the temperature. On the other hand, \citet{Gonj2015} did not find oscillations when considering the more realistic case of non-uniform and anisotropic gaps, in which reactions become activated only in a small fraction of the core volume. We explore this effect for the case of Murca reactions with $\Delta_{\mathrm{thr}}=1.5\,\mathrm{MeV}$ in Fig.~\ref{oscillations}, with reaction rates multiplied by different factors $f\leq 1$. We see that, for $f\gtrsim 10^{-1}$, oscillations occur, whereas for $f\lesssim 10^{-2}$ they are eliminated and the MSP reaches a quasi-steady state that is nearly independent of $f$ (compare the curves with $f=10^{-2}$ and $10^{-5}$). We found the same behavior for Durca reactions with $\Delta_{\mathrm{thr}}=1.5\,\mathrm{MeV}$, in this case with a threshold value $f\sim 10^{-8}$.}

\end{appendix}

\end{document}